\documentclass{article}

\usepackage[utf8]{inputenc}
\usepackage[a4paper, total={6in, 8in}]{geometry}
\usepackage{amsmath}
\usepackage{graphicx}

\title{Radiation Reaction of an accelerating Point Charge: A new Approach}
\author{Nikhil D. Hadap\\Bhabha Atomic Research Center\\Trombay, Mumbai- 400085, India}
\date{\today}

\begin{document}

\maketitle

\begin{abstract}
Abraham Lorentz (AL) formula of Radiation Reaction and its relativistic generalization, Abraham Lorentz Dirac (ALD) formula, are valid only for periodic (accelerated) motion of a charged particle, where the particle returns back to its original state. Thus, they both represent time averaged solutions for radiation reaction force. In this paper, another expression has been derived for radiation reaction following a new approach, starting from Larmor formula, considering instantaneous change (rather than periodic change) in velocity, which is a more realistic situation. Further, it has been also shown that the new expression for Radiation Reaction is free of pathological solutions; which were unpleasant parts of AL as well as ALD equations; and remained unresolved for about 100 years.
\end{abstract}

\section{Introduction:}

One of the most important results of classical electrodynamics is the fact that accelerated (or decelerated) charge radiates away energy in the form of electromagnetic waves. The formula for total power radiated was first derived by J. J. Larmor in 1897; known as the Larmor formula \cite{grif}:

\begin{equation} \label{1}
    P\ =\ \frac{\mu _0q^2a^2}{6\pi c}
\end{equation}

The Larmor formula is applicable for non-relativistic particles; where $v << c$. Relativistic generalization of Larmor formula leads to:

\begin{equation} \label{2}
    P\ =\ \frac{\mu _0q^2\gamma ^6}{6\pi c}\left[a^2-\frac 1{c^2}({\vec v}\times {\vec a})^2 \right]
\end{equation}

Which is Li\'enard's result; which was first obtained in 1898 \cite{grif}. \par

As radiation carries energy and momentum; in order to satisfy energy and momentum conservation, the charged particle must experience a recoil at the time of emission. The radiation must exert an additional force ${\vec F}_{rad}$ back on the charged particle. This force is known as Radiation Reaction Force or simply, Radiation Reaction. \par

In 1902 Max Abraham derived a non- relativistic formula for radiation- reaction by using Larmor formula (\ref{1}), assuming cyclic change in particle's velocity \cite{grif}, by evaluating:

\begin{equation} \label{3}
    \int _{t1}^{t2} {\vec F}_{rad} \cdot {\vec v}\ dt\ =\ -\frac{\mu _0q^2 \gamma ^6}{6\pi c}\int _{t1}^{t2}a^2\ dt
\end{equation}

Which led to the well known Abraham- Lorentz (AL) formula:

\begin{equation} \label{4}
    {\vec F}_{rad}\ =\ \frac{\mu _0q^2}{6\pi c}\frac{d{\vec a}} {dt}
\end{equation}

In 1916, H. A. Lorentz showed that the AL force is actually the ''Self Force'' \cite{lor} which is exerted by the accelerated (extended) charge on itself, in order to resist change in state of motion. \par

In 1938, P. A. M. Dirac derived another formula for radiation reaction; by making use of advanced as well as retarded fields and evaluating stress-energy- tensor for point particles \cite{dir}:

\begin{equation} \label{5}
    F_{{rad}\ \mu }\ =\ \frac{\mu _0q^2}{6\pi mc} \left[ \frac{d^2 p_{\mu }}{d \tau ^2}\ +\frac{p_{\mu }}{m^2 c^2} \left( \frac{d p_{\nu }}{d\tau } \frac{d p^{\nu }}{d\tau } \right) \right]
\end{equation}

Which is, actually, the relativistic generalization of AL formula \cite{roh}; and is known as Abraham- Lorentz- Dirac (ALD) formula. \par

The 3- force (time- space part) of \ref{5} gives:

\begin{equation} \label{5-a}
    {\vec F}_{rad}\ =\ \frac{\mu _0q^2}{6\pi c} \left[ \gamma ^3 \frac{d{\vec a}}{dt}\ + \frac{3 \gamma ^5} c ( {\vec a} \cdot {\vec{\beta }} ) {\vec a} \ +\gamma ^5 ( {\vec{\beta }} \cdot \frac{d{\vec a}}{dt} ) {\vec{\beta }} \ + \frac{3 \gamma ^7} c ( {\vec{\beta }} \cdot {\vec a} ) ^2 {\vec{\beta }} \right]
\end{equation}

The main problem with AL (and also ALD) formula is its run-away solution \cite{grif} \cite{jak}: $a\ =\ a_0e^{t/\tau }$ ; where  $\tau \ =\ (\mu _0q^2)/(6\pi c)$ ; Which says that a charged particle initially at rest, would spontaneously accelerate without bounds. \par

Further, when equation of motion is solved assuming presence of non-zero external force, the solution \cite{jak}:

\begin{equation} \label{6}
    m \frac{d{\vec v}}{dt}\ =\ \frac 1{\tau }\ e^{\frac t{\tau }}\int _t^{\infty} e^{\frac{-t}{\tau}}\ {\vec F}(t')\ {dt'}
\end{equation}

predicts acausal pre-acceleration; i.e. the particle accelerates in advance before the force would act. \par

The AL force is the result of the most fundamental calculation of the effect of self-generated fields. However, why it gives unpleasant pathological solutions, is still an unsolved mystery in classical electrodynamics. \par

It looks the problem may be due to positive sign at RHS; in both (\ref{4}) and (\ref{5}). Dirac also pointed out possible problem of sign in his expression (\ref{5}), in his 1938 paper \cite{dir}. \par

Further, The AL force is the average force that an accelerating charged particle feels in the recoil from the emission of radiation, between the two identical states. Thus, it represents an special case, which requires the particle to return to its initial state of motion, at the end of a cycle. \par

However, a charged particle does radiate while in non periodic accelerated motion; i.e. when it accelerates and finally settles at different velocity. \par

The purpose of this paper to present a new formula for radiation reaction, starting from Larmor formula while considering general case of instantaneously accelerated charge. \par

In section- \ref{derivation}, general expression for radiation- reaction has been derived. \par

In section- \ref{analysis}, a qualitative picture of radiation reaction has been presented. \par

In section- \ref{direction_of_Frad}, direction of ${\vec F}_{rad}$ has been determined to complete final expression for radiation- reaction. \par

In section- \ref{eqn_of_motion}, equation of motion has been solved, to test validity of new expression for ${\vec F}_{rad}$. \par

Section- \ref{conclusion} is the conclusion, followed by acknowledgement and references.

\section{Derivation of a new expression for Radiation Reaction Force:} \label{derivation}

Starting with Larmor- Li\'enard formula (\ref{2}) for electromagnetic power radiated by an accelerating charge; and modifying cross product term by applying scalar triple product rule: ${\vec A} \cdot ({\vec B} \times {\vec C})\ ={\vec B} \cdot ({\vec C} \times {\vec A})$, yields:

\begin{equation} \label{7}
    ({\vec v} \times {\vec a})^2\ =({\vec v} \times {\vec a}) \cdot ({\vec v} \times {\vec a})\ =\ v^2a^2\ -({\vec v} \cdot {\vec a})^2
\end{equation}

Putting the same in Larmor- Li\'enard formula \ref{2} yields:

\begin{equation} \label{8}
    P\ =\ \frac{\mu _0 q^2}{6 \pi c}\ \left[ \gamma^4 a^2\ -\frac{\gamma^6}{c^2}\ ({\vec v} \cdot {\vec a})^2 \right]
\end{equation}

Whenever there is change in charge's velocity $d{\vec v}$ there would be radiation. Thus, any radiation received from a point charge would indicate that there was a change in its velocity at earlier times. Thus, according to conservation of energy, the radiation received (at present time) must equal the negative work- done by Radiation Reaction Force (at retarded time):

\begin{equation} \label{9}
    P\ =\ -\int _{v1}^{v2} {\vec F}_{rad} \cdot {\vec{dv}}\ =\ -\int _{t1}^{t2} {\vec F}_{rad} \cdot {\vec a}\ {dt}
\end{equation}

which implies:

\begin{equation} \label{10}
    {\vec F}_{rad} \cdot {\vec a}\ =\ -\frac{dP}{dt}\ =\frac{-\mu_0 q^2}{6 \pi c}\ \frac{d}{dt} \left[ \gamma^4 a^2\ -\frac{\gamma ^6}{c^2} ( {\vec v} \cdot {\vec a})^2 \right]
\end{equation}

Where $t$ is retarded coordinate time. Carrying out differentiation by parts yields:

\begin{equation} \label{10-a}
    {\vec F}_{rad}\ \cdot {\vec a}\ =\ -\frac{\mu _0 q^2}{3 \pi c}\ \left[ \gamma^4 ( {\vec a} \cdot \frac{d{\vec a}}{dt})\ +\frac{3 \gamma^6}{c^2}( {\vec v} \cdot {\vec a})\ a^2\ +\frac{\gamma^6}{c^2}( {\vec v} \cdot {\vec a})( {\vec v} \cdot \frac{d{\vec a}}{dt} )\ +\frac{3 \gamma^6}{c^4}( {\vec v} \cdot {\vec a} )^3 \right]
\end{equation}

Or:

\begin{equation} \label{10-b}
    {\vec F}_{rad}\ \cdot {\vec a}\ =\ -\frac{\mu _0 q^2}{3\pi c}\ \left[ \gamma^4 ( {\vec a} \cdot \frac{d{\vec a}}{dt} )\ +\frac{3 \gamma ^6}{c} ( {\vec{\beta}} \cdot {\vec a} )a^2\ +\gamma ^6 ( {\vec{\beta }} \cdot {\vec a} ) ( {\vec{\beta }} \cdot \frac{d{\vec a}}{dt} )\ +\frac{3 \gamma ^8}{c} ( {\vec{\beta }} \cdot {\vec a})^3 \right]
\end{equation}

As equation (\ref{10-b}) is an scalar expression, it doesn't represent unique solution for ${\vec F}_{rad}$. However, it does have a general solution:

\begin{equation} \label{11} 
\begin{split}
      {\vec F}_{rad}\ =\ -\frac{\mu_0 q^2}{3 \pi c} \left[ \gamma^4 \frac{d{\vec a}}{dt}\ +\frac{3 \gamma^6}{c} x({\vec a} \cdot {\vec{\beta}}) {\vec a}\ +\frac{3 \gamma^6}{c} (1-x)a^2 {\vec{\beta}}\ +\gamma^6 ( {\vec{\beta}} \cdot {\vec{\frac{d{\vec a}}{dt}}} ) {\vec{\beta}} 
      \right. \\ \left. 
      +\frac{3\gamma^8}{c} ( {\vec{\beta}} \cdot {\vec a} )^2 {\vec{\beta }} \right]
\end{split} 
\end{equation}

Where, $x$ is a number; and 0 ${\leq}$ x ${\leq}$ 1. \par

Contrary to AL equation (\ref{4}) and ALD equation (\ref{5}), here in (\ref{11}), there is negative sign at RHS. Thus, the expression is expected to be free from runaway solutions. \par

However, (\ref{11}) is not a complete expression; because the value of $x$ is yet to be determined; which shall be evaluated later on, in this paper. \par

However, before proceeding further, a qualitative analysis of radiation reaction has been presented in the next section, which would provide valuable information about ${\vec F}_{rad}$. \par

\section{Qualitative Picture of Radiation Reaction:} \label{analysis}

Rewriting the obtained general expression of radiation reaction (\ref{11}) in vector component form:

\begin{equation} \label{11-a} 
\begin{split}
      {\vec F}_{rad}\ =\ -\frac{\mu_0 q^2}{3 \pi c} \left[ \gamma^4 \frac{d{\vec a}}{dt}\ +\frac{3 \gamma^6}{c} x({\vec a} \cdot {\vec{\beta}}) {\vec a}\ + \left\{\  \frac{3 \gamma^6}{c} (1-x)a^2\ +\gamma^6 ( {\vec{\beta}} \cdot {\frac{d{\vec a}}{dt}} )  
      \right. \right. \\ \left. \left.
      +\frac{3\gamma^8}{c} ( {\vec{\beta}} \cdot {\vec a} )^2 \right\} {\vec{\beta }}\ \right]
\end{split} 
\end{equation}

The first term of (\ref{11-a}) represents component of ${\vec F}_{rad}$ that points opposite to the instantaneous direction of jerk (direction of change in acceleration). This term is like Schott term of ALD equation (\ref{5}); the only term which survives in particle's rest frame ($\beta=0$). Effectively, this term appears to resist any change in acceleration. \par

The second term of (\ref{11-a}) represents component of ${\vec F}_{rad}$ that points opposite to the instantaneous direction of acceleration. Effectively, this term appears to resist acceleration; i.e. the change in velocity. \par

Remaining terms of (\ref{11-a}) comprise Radiation Recoil term, which points opposite to the instantaneous direction of velocity. Magnitude of which depends upon instantaneous values of jerk and acceleration, coupled with instantaneous velocity. \par

The derived expression (\ref{11-a}) represents instantaneous values of ${\vec F}_{rad}$. However, ${\vec F}_{rad}$ is still a ''space-averaged'' quantity; because it has been derived from Larmor- Li\'enard formula (\ref{2}), which is ''Total Power radiated in All Directions''.

Thus, at any instance any ''Radiation Emitted in all Space'' is a cause; and radiation reaction is its effect; its ''Mirror Image''.

Therefore, valuable information about direction of ${\vec F}_{rad}$ can be inferred from examination of shape of radiation patterns; which must always point opposite to the ''Net Direction of radiation''.

\begin{figure} 
    \centering
    \includegraphics{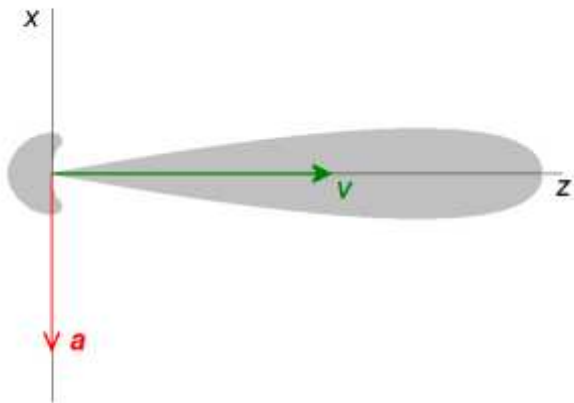}
    \caption{(Color Online) Radiation Pattern emitted by a charge under Circular motion}
    \label{fig:2}
\end{figure}

Figure \ref{fig:1} and Figure \ref{fig:2} are two specific cases of radiation profiles from accelerated charge particles:

\begin{figure} 
    \centering
    \includegraphics{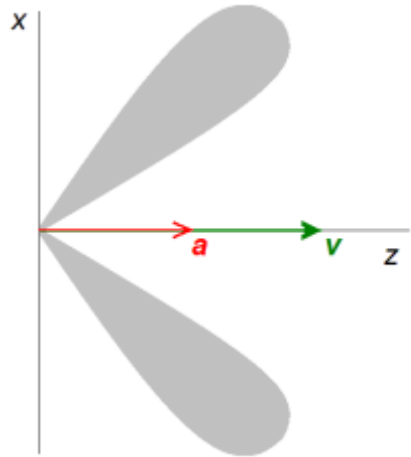}
    \caption{(Color Online) Radiation Pattern emitted by a Linearly accelerated charge}
    \label{fig:1}
\end{figure}

Figure \ref{fig:1} represents radiation patterns from a linearly accelerated charge (${\vec a}\parallel{\vec v}$), as in case of Bremsstrrahlung. Whereas, figure \ref{fig:2} represents radiation from a charge under circular motion (${\vec a}\perp {\vec v}$ ), as in case of Synchrotron radiation. \par

In both the cases, radiation patterns are symmetrical about velocity vector. Thus, they provide an idea that ${\vec F}_{rad}$ must point in $- \vec v$ direction; and therefore suggest $x = 0$ in equation (\ref{11}). \par

\begin{figure} 
    \centering
    \includegraphics[width=12cm]{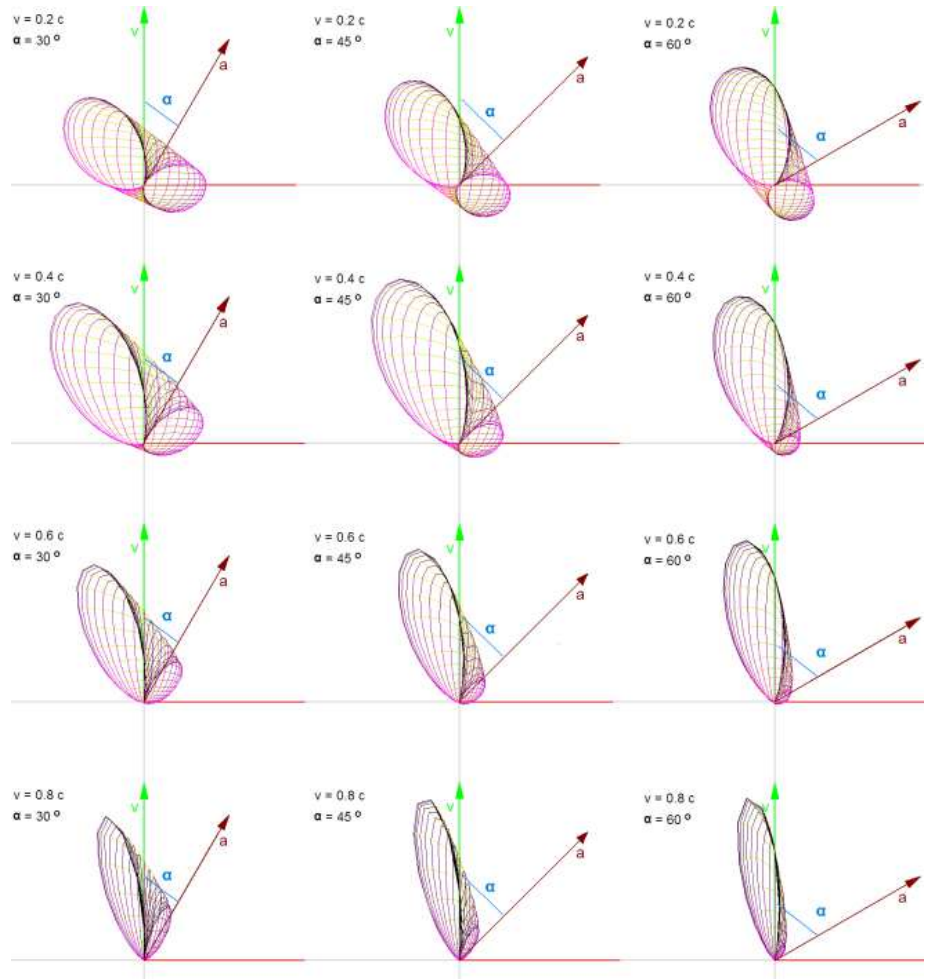}
    \caption{(Color Online) Snapshots of Radiation Patterns of a charged particle with acceleration vectors pulling its trajectory at various angles.}
    \label{fig:3}
\end{figure}

However, in general case when ${\vec a}$ and ${\vec v}$ were at certain angle $\alpha$, and $ 0< \alpha< \pi $, shape of radiation patterns Figure \ref{fig:3} provide no help in deducing direction of ${\vec F}_{rad}$ directly. \par

Therefore, direction of ${\vec F}_{rad}$ needs to be found analytically, which has been presented in the next section.

\section{Direction of Radiation Reaction Vector:} \label{direction_of_Frad}

Starting with the Poyenting vector for radiation fields, given by \cite{grif}:

\begin{equation} \label{12}
    {\vec S}_{{RAD}}\ =\ \left( \frac{ {\vec R} \cdot {\vec U} }{Rc} \right) \frac{1}{\mu_0 c} E_{RAD}^2 {\hat n}\ =\ ( 1-{\hat n} \cdot {\vec{\beta }} ) \frac{1}{\mu_0 c} E_{RAD}^2 {\hat n}
\end{equation}

Where: \par

${\hat n}=\ {\vec R}/R$, the unit vector in the direction of observer, at large distance R, \par

${\vec U}=\ c{\hat n}\ -{\vec v}$, the Retarded relative vector, \& \par

${\vec E}_{RAD}$ is Radiation Electric Field; given by \cite{grif}:

\begin{equation} \label{13}
\begin{split}
    {\vec E}_{RAD}\ =\ \frac{q}{4 \pi \epsilon_0} \frac{R}{ ({\vec R} \cdot {\vec U})^3 } \left\{ {\vec R} \times  ({\vec U} \times {\vec a}) \right\}\ =\ \frac{q}{4 \pi \epsilon_0 R c^2} \left\{ \frac{ ({\hat n} \cdot {\vec a}) }{(1-{\hat n} \cdot {\vec{\beta }})^3} 
    \right. \\ \left. 
    -\frac{{\vec a}}{(1- {\hat n} \cdot {\vec{\beta }})^2} \right\}    
\end{split}
\end{equation}

Thus, using (\ref{13}) in (\ref{12}) the expression for Radiation Poyenting Vector becomes:

\begin{equation} \label{12-a}
    {\vec S}_{RAD}\ =\ \frac{q^2 {\hat n}}{16 \pi^2 \epsilon_0 R^2 c^3} \left\{ \frac{a^2}{ (1-{\hat n} \cdot {\vec{\beta}})^3 }\ +\frac{2 ({\hat n} \cdot {\vec a})({\vec a} \cdot {\vec{\beta }}) }{(1-{\hat n} \cdot {\vec{\beta }})^4}\ -\frac{(1-\beta^2)({\hat n} \cdot {\vec a})^2}{(1-{\hat n} \cdot {\vec{\beta}})^5} \right\}
\end{equation}

Using Cartesian coordinate vectors:

${\vec{\beta}}\ =\ \beta{\hat z}$, \par

${\vec a}\ =\ a(\sin \alpha {\hat x}\ +cos \alpha {\hat z})$, \& \par

${\hat n}\ =\ \sin \theta \cos \phi {\hat x}\ +sin \theta \sin \phi {\hat y}\ +cos \theta {\hat z}$ \par

where, $\alpha$ is the angle between particle's instantaneous acceleration and instantaneous velocity; the equation (\ref{12-a}) becomes:

\begin{equation} \label{12-b}
\begin{split}
    {\vec S}_{RAD}\ =\ \frac{q^2 a^2}{16 \pi^2 \epsilon_0 R^2 c^3} \left[ \frac{1}{(1-\beta \cos \theta )^3}\ +\frac{2 \beta \sin \alpha \cos \alpha \sin \theta \cos \phi }{(1-\beta \cos \theta)^4}\ +\frac{2 \beta \cos^2 \alpha \cos \theta}{(1-\beta \cos \theta )^4}
    \right. \\ \left.
    -\frac{(1-\beta^2) \sin^2 \alpha \sin^2 \theta \cos^2 \phi }{(1-\beta \cos \theta)^5}\ -\frac{(1-\beta^2) \cos^2 \alpha \cos^2 \theta}{(1-\beta \cos \theta )^5} 
    \right.\\ \left.
    -\frac{2 (1-\beta^2) \sin \alpha \cos \alpha \sin \theta \cos \theta \cos \phi}{(1-\beta \cos \theta)^5} \right] (\sin \theta \cos \phi {\hat x}\ +\sin \theta \sin \phi {\hat y}\ +\cos \theta {\hat z})    
\end{split}
\end{equation}

\begin{figure} 
    \centering
    \includegraphics{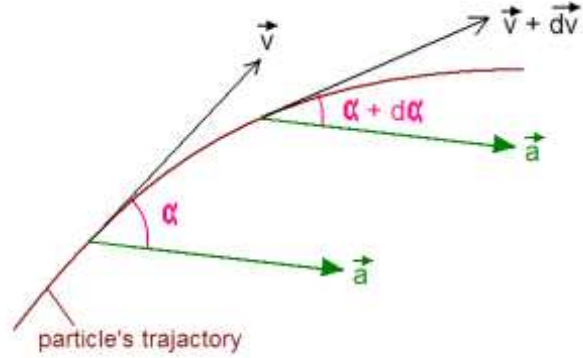}
    \caption{(Color Online) Particle's Velocity \& Acceleration vectors in space, along its trajectory.}
    \label{fig:4}
\end{figure}

However, $\alpha=0$ cannot remain constant (unless $\alpha=0$ or $\pi/ 2$); because, component of $\vec a$ perpendicular to $\vec v$ would continuously pull particle's trajectory more and more towards $\vec a$ (\ref{fig:4}). Thus $\alpha$ would continuously change in time. \par

\begin{figure} 
    \centering
    \includegraphics{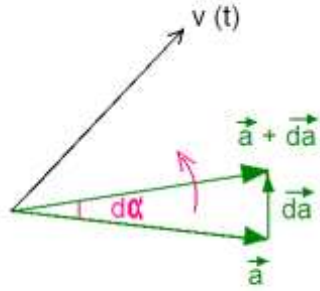}
    \caption{(Color Online) Velocity \& Acceleration vectors in particle's own Rest Frame.}
    \label{fig:5}
\end{figure}

Even if, the acceleration (or applied force) were constant in space; in particle's frame it would seem to be changing in the direction of changing $\alpha$. The direction of change in $\alpha$ would be the same as the the direction of $d{\vec a}/{dt}$ (\ref{fig:5}). \par

This is why $d{\vec a}/{dt}$ term comes into the expression of ${\vec F}_{rad}$; which is a higher order term, and becomes significant when history of particle's motion (over a certain small period) were also taken into account. \par

However, purpose of this analysis is to know components of ${\vec F}_{rad}$ along $\vec a$ and $\vec v$ (to know value of $x$ in equation (\ref{11})). So only ''instantaneous snapshot'' of particle's state of motion  is being considered.

At any instance radiation reaction acts like ''Mirror Image'' to the ''Radiation Emitted in all Space''. Therefore, if negative of all radiation poyenting vectors (\ref{12-b}) were algebraically added up together, over all space, the resultant vector ${\vec W}$ would point in the direction of ${\vec F}_{rad}$:

\begin{equation} \label{15}
     {\vec W}\ =\ -\oint_v {\vec S}_{RAD}\ {da}\ =\ -\int_{\theta=0}^{\pi} \int_{\phi=0}^{2 \pi} {\vec S}_{RAD}\ R^2 \sin \theta\ d\theta d\phi
\end{equation}

Substituting expression of ${\vec S}_{RAD}$ from (\ref{12-b}) and writing only non- vanishing integrals in $\phi$ yields:

\begin{equation} \label{15-a}
\begin{split}
    {\vec W}\ =\ -\frac{q^2 a^2}{16 \pi^2 \epsilon_0 c^3} \left[ {\hat x} \sin \alpha \cos \alpha \left\{ 2 \beta \int_{\theta=0}^{\pi} \int_{\phi=0}^{2 \pi} \frac{\sin^3 \theta \cos^2 \phi \ d\theta d\phi}{(1-\beta \cos \theta )^4} 
    \right. \right. \\ \left. \left.
    -2(1-\beta^2) \int_{\theta=0}^{\pi}\int_{\phi=0}^{2 \pi} \frac{\sin^3 \theta \cos \theta \cos^2 \phi \ d\theta d\phi }{(1-\beta \cos \theta)^5} \right\} 
    \right. \\ \left.
    +{\hat z} \left\{ \int_{\theta=0}^{\pi} \int_{\phi=0}^{2 \pi} \frac{\sin \theta \cos \theta \ d\theta d\phi }{(1-\beta \cos \theta )^3}\ +2\beta \cos^2 \alpha \int_{\theta=0}^{\pi} \int_{\phi=0}^{2 \pi} \frac{\sin \theta \cos^2 \theta \ d\theta d\phi}{(1-\beta \cos \theta)^4} 
    \right. \right. \\ \left. \left.
    -(1-\beta^2) \sin^2 \alpha \int_{\theta=0}^{\pi} \int_{\phi=0}^{2 \pi} \frac{\sin^3 \theta \cos \theta \cos^2 \phi \ d\theta d\phi}{(1-\beta \cos \theta)^5}
    \right. \right. \\ \left. \left.
    -(1-\beta^2) \cos^2 \alpha \int_{\theta=0}^{\pi} \int_{\phi=0}^{2 \pi} \frac{\sin \theta \cos^3 \theta \ d\theta d\phi}{(1-\beta \cos \theta)^5} \right\} \right]
\end{split}
\end{equation}

Substituting suitable  $\phi$ integral values yields:

\begin{equation} \label{15-b}
\begin{split}
      {\vec W}\ =\ \frac{-q^2 a^2}{16 \pi^2 \epsilon_0 c^3} \left[ {\hat x} \sin \alpha \cos \alpha \left\{ 2 \pi \beta \int_{0}^{\pi} \frac{\sin^3 \theta \ d\theta}{(1-\beta \cos \theta )^4}\ -2 \pi (1-\beta ^2) \int_{0}^{\pi} \frac{\sin^3 \theta \cos \theta \ d\theta }{(1-\beta \cos \theta)^5} \right\} 
      \right. \\ \left. 
      +{\hat z} \left\{ 2 \pi \int_{0}^{\pi} \frac{\sin \theta \cos \theta \ d\theta}{(1-\beta \cos \theta )^3}\ +4 \pi \beta \cos^2 \alpha \int_{0}^{\pi} \frac{\sin \theta \cos^2 \theta \ d\theta}{(1-\beta \cos \theta)^4}
      \right. \right. \\ \left. \left.
      -\pi (1-\beta ^2) \sin^2 \alpha \int_{0}^{\pi} \frac{\sin^3 \theta \cos \theta \ d\theta}{(1-\beta \cos \theta)^5}\ -2 \pi (1-\beta^2) \cos^2 \alpha \int _{0}^{\pi} \frac{\sin \theta \cos^3 \theta \ d\theta}{(1-\beta \cos \theta)^5} \right\} \right]
\end{split}
\end{equation}

Rearranging the $\sin \theta$ and $\cos \theta$ terms; (\ref{15-b}) simplifies to:

\begin{equation} \label{15-c}
\begin{split}
    {\vec W}\ =\ -\frac{q^2 a^2}{16 \pi \epsilon_0 c^3} \left[ 2 {\hat x} \sin \alpha \cos \alpha \left\{ \beta \int_{0}^{\pi} \frac{\sin \theta \ d\theta}{(1-\beta \cos \theta)^4}\ -\beta \int_{0}^{\pi} \frac{\sin \theta \cos^2 \theta \ d\theta}{(1-\beta \cos \theta)^4}
    \right. \right. \\ \left. \left.
    -(1-\beta^2) \int_{0}^{\pi} \frac{\sin \theta \cos \theta \ d\theta}{(1-\beta \cos \theta)^5}\ +(1-\beta^2) \int_{0}^{\pi} \frac{\sin \theta \cos^3 \theta \ d\theta }{(1-\beta \cos \theta)^5} \right\}
    \right. \\ \left.
    +{\hat z} \left\{ 2 \int_{0}^{\pi} \frac{\sin \theta \cos \theta \ d\theta}{(1-\beta \cos \theta)^3}\ +4\beta \cos ^2\alpha \int_{0}^{\pi} \frac{\sin \theta \cos^2 \theta \ d\theta}{(1-\beta \cos \theta)^4}
    \right. \right. \\ \left. \left.
    -(1-\beta^2) \sin^2 \alpha \int_{0}^{\pi} \frac{\sin \theta \cos \theta \ d\theta}{(1-\beta \cos \theta)^5}\ +(1-\beta^2) \sin^2 \alpha \int_{0}^{\pi} \frac{\sin \theta \cos^3 \theta \ d\theta}{(1-\beta \cos \theta)^5}
    \right. \right. \\ \left. \left.
    -2(1-\beta^2) \cos^2 \alpha \int_{0}^{\pi} \frac{\sin \theta \cos^3 \theta \ d\theta}{(1-\beta \cos \theta)^5} \right\} \right]
\end{split}
\end{equation}

Solving the integrals by substitution  $\beta \cos \theta= t$; $\sin \theta d\theta= -dt/ \beta$ and changing the limits ($\theta= 0$ to  $\pi$) to ($t= \beta$ to  $-\beta$ ) yields:

\begin{equation} \label{15-d}
\begin{split}
    {\vec W}\ =\ \frac{q^2 a^2}{16 \pi \epsilon_0 c^3} \left[ 2 {\hat x} \sin \alpha \cos \alpha \left\{ \int_{\beta }^{-\beta } \frac{dt}{(1-t)^4}\ -\frac{1}{\beta^2} \int_{\beta }^{-\beta} \frac{t^2\ dt}{(1-t)^4}\ -\frac{(1-\beta ^2)}{\beta ^2} \int_{\beta}^{-\beta} \frac{t\ dt}{(1-t)^5}
    \right. \right. \\ \left. \left.
    +\frac{(1-\beta^2)}{\beta^4} \int_{\beta}^{-\beta} \frac{t^3\ dt}{(1-t)^5} \right\}\ + {\hat z} \left\{ \frac{2}{\beta^2} \int_{\beta}^{-\beta} \frac{t\ dt}{(1-t)^3}\ +\frac{4}{\beta^2} \cos^2 \alpha \int_{\beta}^{-\beta} \frac{t^2\ dt}{(1-t)^4}
    \right. \right. \\ \left. \left.
    -\frac{(1-\beta^2)}{\beta^2} \sin^2 \alpha \int_{\beta}^{-\beta} \frac{t\ dt}{(1-t)^5}\ +\frac{(1-\beta^2)}{\beta^4} \sin^2 \alpha \int_{\beta}^{-\beta} \frac{t^3\ dt}{(1-t)^5}
    \right. \right. \\ \left. \left.
    -2 \frac{(1-\beta^2)}{\beta^4} \cos^2 \alpha \int_{\beta}^{-\beta} \frac{t^3\ dt}{(1-t)^5} \right\} \right]
\end{split}
\end{equation}

The integrals can be solved easily; either by parts or by partial fraction. Following are the solutions:

\begin{equation} \label{16}
\begin{split}
    \int_{\beta}^{-\beta} \frac{t\ dt}{(1-t)^3}\ =\ -\frac{2 \beta^3}{(1-\beta^2)^2};\\
    \int_{\beta}^{-\beta} \frac{dt}{(1-t)^4}\ =\ -\frac{2 \beta}{3} \frac{(3+ \beta^2)}{(1-\beta^2)^3};\\
    \int_{\beta}^{-\beta} \frac{t^2\ dt}{(1-t)^4}\ =\ -\frac{2\beta^3}{3} \frac{(1+3\beta^2)}{(1-\beta ^2)^3};\\
    \int_{\beta}^{-\beta} \frac{t\ dt}{(1-t)^5}\ =\ -\frac{2 \beta^3}{3} \frac{(5+\beta^2)}{(1-\beta^2)^4}\\
    \int_{\beta}^{-\beta} \frac{t^3\ dt}{(1-t)^5}\ =\ -2 \beta^5 \frac{(1+\beta ^2)}{(1-\beta ^2)^4}
\end{split}
\end{equation}

Putting the values of various integrals (\ref{16}) into (\ref{15-d}) yields:

\begin{equation} \label{15-e}
\begin{split}
    {\vec W}\ =\ \frac{q^2 a^2 \beta}{16 \pi \epsilon_0 c^3\ (1-\beta^2)^3} \left[ 2 {\hat x} \sin \alpha \cos \alpha \left\{ \frac{-2}{3}(3+\beta^2)\ +\frac{2}{3}(1+3\beta^2)\ +\frac{2}{3}(5+\beta^2)
    \right. \right. \\ \left. \left.
    -2(1+\beta^2) \right\}\ +{\hat z} \left\{ -4 (1-\beta ^2)\ -\frac{8}{3} (1+3\beta^2) \cos^2 \alpha
    \right. \right. \\ \left. \left.
    +\frac{2}{3} (5+\beta^2) \sin^2 \alpha \ -2 (1+\beta^2) \sin^2 \alpha \ +4 (1+\beta^2) \cos^2 \alpha \right\} \right]
\end{split}
\end{equation}

Which gets simplified to:

\begin{equation} \label{15-f}
    {\vec W}\ =\ -\frac{q^2 a^2 \beta \gamma^6}{6 \pi \epsilon_0 c^3}\ {\hat z} \cos^2 \alpha
\end{equation}

which points in $-{\hat z}$ direction, i.e. opposite to the direction of instantaneous velocity (like friction force), for any value of $\alpha$. \par

However, if ${\vec F}_{rad}$ indeed points in $-{\vec v}$ direction, its expression must not contain any term going with ${\vec a}$. Thus, the correct expression for ${\vec F}_{rad}$ is equation (\ref{11-a}) with $x= 0$. \par

Rewriting the final expression for Radiation Reaction vector:

\begin{equation} \label{11-d}
    {\vec F}_{rad}\ =\ -\frac{\mu_0 q^2}{3 \pi c} \left[ \gamma^4 \frac{\vec da}{dt}\ +\left\{ \frac{3 \gamma^6}{c} a^2\ +\gamma^6 (\vec{\beta} \cdot \frac{\vec da}{dt})\ +\frac{3 \gamma^8}{c} ({\vec{\beta}} \cdot {\vec a})^2 \right\} {\vec{\beta }} \right]
\end{equation}

Further, equation (\ref{11-d}) can be expressed in terms of relativistic mechanical force (or rate of change of momentum) which is experienced by the charged particle:

\begin{equation} \label{17}
    {\vec F}_{MECH}\ =\ \frac{\vec dp}{dt}\ =\ m \frac{d(\gamma {\vec v})}{dt}\ =\ \gamma m {\vec a}\ +\frac{\gamma^3 m}{c^2}({\vec v} \cdot {\vec a}) {\vec v}
\end{equation}

Differentiating (\ref{17}), once again w.r.t. (retarded) time, yields:

\begin{equation} \label{18}
\begin{split}
    \frac{{\vec dF}_{MECH}}{dt}\ =\ \frac{d^2 {\vec p}}{{dt}^2}\ =\ m \gamma \frac{\vec da}{dt}\ +m \frac{d \gamma}{dt} {\vec a}\ +\frac{m \gamma^3}{c^2}({\vec v} \cdot {\vec a}) {\vec a}\ +\frac{m \gamma^3}{c^2}({\vec v} \cdot \frac{\vec da}{dt}) {\vec v} \\ 
    +\frac{m \gamma^3}{c^2} a^2 {\vec v}\ +\frac{3 m \gamma^2}{c^2}({\vec v} \cdot {\vec a}) \frac{d \gamma}{dt} {\vec v}
\end{split}
\end{equation}

Meanwhile:

\begin{equation} \label{19}
    \frac{d\gamma}{dt}\ =\ \frac{\gamma^3}{c^2}({\vec v} \cdot {\vec a})
\end{equation}

Thus, it yields from equation (\ref{18}):

\begin{equation} \label{20}
\begin{split}
    \frac{1}{m} \frac{{\vec dF}_{MECH}}{dt}\ =\ \frac{1}{m} \frac{d^2 {\vec p}}{{dt}^2}\ =\ \gamma \frac{\vec da}{dt}\ +\frac{2 \gamma^3}{c}({\vec{\beta}} \cdot {\vec a}) {\vec a}\ +\left\{ \gamma^3 \frac{a^2}{c}\ +\frac{3 \gamma^5}{c} ({\vec{\beta}} \cdot {\vec a})^2 
    \right. \\ \left.
    +\gamma^3 \left( {\vec{\beta}} \cdot \frac{\vec da}{dt} \right) {\vec{\beta }} \right\} {\vec{\beta }}
\end{split}
\end{equation}

Or:

\begin{equation} \label{20-a}
\begin{split}
    \frac{1}{m} \frac{{\vec dF}_{MECH}}{dt}\ =\ \frac{1}{m} \frac{d^2{\vec p}}{{dt}^2}\ =\ \gamma \frac{\vec da}{dt}\ +\left\{\frac{3 \gamma ^3 a^2}{c}\ +\frac{3 \gamma^5}{c} ({\vec{\beta}} \cdot {\vec a})^2\ +\gamma^3 \left( {\vec{\beta}} \cdot \frac{\vec da}{dt} \right) {\vec{\beta}} \right\} {\vec{\beta}} \\ 
    +\frac{2 \gamma^3}{c} ({\vec{\beta }} \cdot {\vec a}) {\vec a}\ -\frac{2 \gamma^3}{c} a^2 {\vec{\beta }}
\end{split}
\end{equation}

Comparing equation (\ref{20-a}) with (\ref{11-d}); there arrives another expression for radiation reaction force:

\begin{equation} \label{11-e}
    {\vec F}_{rad}\ =\ -\frac{\mu_0 q^2}{3 \pi c} \left[ \frac{\gamma^3}{m} \frac{{\vec dF}_{MECH}}{dt}\ -\frac{2 \gamma^6}{c} \{ {\vec a} \times ({\vec a} \times {\vec{\beta}}) \} \right]
\end{equation}

Or:

\begin{equation} \label{11-f}
    {\vec F}_{rad}\ =\ -\frac{\mu_0 q^2}{3 \pi c} \left[ \frac{\gamma^3}{m} \frac{d^2 {\vec p}}{{dt}^2}\ -\frac{2 \gamma^6}{c} \{ {\vec a} \times ({\vec a} \times {\vec{\beta}}) \} \right]
\end{equation}

\section{Equation of Motion:} \label{eqn_of_motion}

Considering non-relativistic case where $\gamma \approx 1$. From Newton's second law, the equation of motion for a charged particle is:

\begin{equation} \label{21}
    {\vec F}\ +{\vec F}_{rad}\ =\ m {\vec a}
\end{equation}

Further, considering motion in straight line (${\vec a} \parallel {\vec{\beta }}$); the second term in equation (\ref{11-f}) drops and the equation of radiation reaction reduces to:

\begin{equation} \label{22}
    {\vec F}_{rad}\ =\ -\frac{\mu_0 q^2 \gamma^3}{3 \pi m c} \frac{d^2 p}{{dt}^2} {\hat x}\ \approx \ -\frac{\mu_0 q^2}{3 \pi c} \frac{da}{dt} {\hat x}
\end{equation}

Using expression of ${\vec F}_{rad}$ from equation (\ref{22}) in equation of motion (\ref{21}) yields:

\begin{equation} \label{22-a}
    F\ {\hat x}\ -\frac{\mu _0 q^2}{3 \pi c} \frac{da}{dt} {\hat x}\ =\ m a\ {\hat x}
\end{equation}

Or:

\begin{equation} \label{22-b}
    \frac{F}{m}\ =\ a\ +\frac{\mu_0 q^2}{3 \pi m c} \frac{da}{dt}\ =\ a\ +\tau \frac{da}{dt}
\end{equation}

Or:

\begin{equation} \label{22-c}
    \frac{F}{\tau\ m}\ =\ \frac{a}{\tau}\ +\frac{da}{dt}
\end{equation}

Where; $\tau =\ \frac{\mu_0 q^2}{3 \pi m c}$; the time constant of a moving charged particle. \par

Using integrating factor $e^{t/\tau}$, the solution of equation (\ref{22-c}) comes out to be:

\begin{equation} \label{23}
    a\ =\ \frac{1}{\tau\ m}\ e^{-t/\tau} \int_{T_0}^t F(t')\ e^{t'/\tau}\ {dt'}
\end{equation}

Which is similar to equation (\ref{6}) but with opposite signs in time variable. The constant of integration $T_0$ needs to be found from physical ground. \par

Here, unlike usual mechanical systems, the acceleration doesn't depend upon instantaneous value of force; but on its weighted time average. Presence of $t'/\tau$ indicate that only small interval of the order of $\tau$ is involved. \par

Integrating by parts in equation (\ref{23}) yields:

\begin{equation} \label{23-a}
    a\ =\ \frac{1}{m} \left\{ F(t)\ -F(T_0)\ e^{(T_0\ -t)/\tau } \right\} \ -\frac{1}{m}\ e^{-t/\tau } \int_{T_0}^t \left\{ \frac{dF}{dt}\ e^{t'/\tau } \right\} {dt}'
\end{equation}

Integrating further, the second term of (\ref{23-a}) by parts yields:

\begin{equation} \label{23-b}
\begin{split}
    a\ =\ \frac{1}{m} \left\{ F(t)\ -F(T_0)\ e^{(T_0\ -t)/\tau } \right\} -\frac{\tau}{m} \left\{ \frac{dF}{dt}(t)\ -\frac{dF}{dt}(T_0)\ e^{(T_0\ -t)/\tau} \right\} \\ +\frac{\tau}{m}\ e^{-t/\tau } \int_{T_0}^t \left\{ \frac{d^2 F}{{dt}^2}\ e^{t'/\tau} \right\} {dt}'
\end{split}
\end{equation}

As $q \to 0$ (for neutral particle); $\tau \to 0$; the acceleration must reduce to $F(t)/ m$, as in case of usual mechanical systems. Thus, (\ref{23-b}) demands $T_0 \leq 0$. \par

Therefore, setting time origin $T_0=0$; the successive integration by parts in (\ref{23-b}) yields:

\begin{equation} \label{23-c}
\begin{split}
    a\ =\ \frac{1}{m} \left\{ F(t)\ -\tau \frac{dF}{dt}(t)\ +\tau^2 \frac{d^2 F}{{dt}^2}\ -\tau^3 \frac{d^3 F}{{dt}^3}\ +.... \right\} \\
    -\frac{e^{-t/\tau }}{m} \left\{ F(0)\ -\tau \frac{dF}{dt}(0)\ +\tau^2 \frac{d^2 F}{{dt}^2}(0)\ -\tau^3 \frac{d^3 F}{{dt}^3}(0)\ +.... \right\}
\end{split}
\end{equation}

Which (using Taylor Series) reduces to:

\begin{equation} \label{23-d}
    a\ =\ \frac{1}{m} \left[ F(t\ -\tau)\ -F(-\tau)\ e^{-t/\tau } \right]
\end{equation}

Thus, there is no pre-acceleration seen here. \par

The second term in (\ref{23-d}) would vanish as $t \to \infty$; i.e. if motion were observed for a long enough time. \par

The first term simply indicates, ``Delay in Response''; i.e. the present acceleration would depend upon force which was received by the particle a little earlier. Thus, causality is not violated here. However, as $\tau $ is too small (of the order of $10^{-24}s$ for fundamental particles), it doesn't come into picture in macroscopic systems. \par

For constant force, the acceleration becomes:

\begin{equation} \label{23-e}
    a\ =\ \frac{F}{m} \left(1\ -e^{-t/\tau } \right)
\end{equation}

Which reminds of equations of current/ voltage growths in LR/ RC circuits respectively; or any first order equation of motion with damping force. \par

Again going back to equation (\ref{22-c}) and considering absence of any external force, $F=0$; the solution comes out to be:

\begin{equation}
    a\ =\ a_0\ e^{-t/\tau }
\end{equation}

Which says that any perturbation in motion of charged particle would die out quickly and the charge would acquire another steady state. Thus here, there is no runaway seen here as well.

\section{Conclusion:} \label{conclusion}

Equations of motion (\ref{23-d}) and (\ref{23-e}), based on newly derived equation of Radiation Reaction, give clear impression that the Radiation Reaction is Electromagnetic Friction force that provides damping effect to the motion of charged particle. \par

The new expressions of Radiation Reaction, (\ref{11-d}) and (\ref{11-f}), successfully eliminate pathological solutions of runaway and acausal pre-accelaration, which were disturbing parts of AL equation (\ref{4}) and ALD equation (\ref{5}) so far. \par

Further, they are found to be in same form and containing the similar terms as in ALD formula (\ref{5-a}). Therefore, they can be believed upon. \par

However, here particle's time constant has been found to be $\mu _0 q^2/(3 \pi mc)$, which is twice that of AL formula (\ref{4}) as well as ALD formula (\ref{5}). The source of this discrepancy might be due to the fact that AL equation (\ref{4}) was derived considering cyclic change in velocity; the process which might have actually averaged out contributions from rising and returning accelerations.

The value of time constant $\tau$ mismatches with Lorentz's theory of self- force \cite{lor} as well. Thus, self force phenomenon needs to be revisited and to be studied further to check this discrepancy.

\section{Acknowledgement}

I am thankful to my institution Bhabha Atomic Research Center that inspired me to learn; and also for providing me facilities and resources that helped me in conceptualizing and writing this paper.

\end{document}